\documentclass[acmsmall]{acmart}

\AtBeginDocument{%
  \providecommand\BibTeX{{%
    \normalfont B\kern-0.5em{\scshape i\kern-0.25em b}\kern-0.8em\TeX}}}

\setcopyright{acmcopyright}
\copyrightyear{0000}
\acmYear{0000}
\acmDOI{xx.xxxx/xxxxxxx.xxxxxxx}

\acmJournal{JACM}
\acmVolume{00}
\acmNumber{00}
\acmArticle{00}
\acmMonth{00}

\usepackage{enumitem}
\begin{document}

\title{An Online Survey on the Perception of Mediated Social Touch Interaction and Device Design}

\author{Carine Rognon}
\email{carinerognon@fb.com}
\orcid{0001-7527-1479}

\author{Taylor~Bunge}
\email{taylorbunge@fb.com}

\author{Meiyuzi~Gao}
\email{gaomei@fb.com}

\author{Chip~Conor}
\email{cconnor@fb.com}

\author{Benjamin~Stephens-Fripp}
\email{bensfripp@fb.com}

\author{Casey~Brown}
\email{casey.brown@fb.com}

\author{Ali~Israr}
\email{aliisrar@fb.com}
\affiliation{%
  \institution{Facebook Reality Labs Research}
  \city{Redmond}
  \state{Washington}
  \country{USA}
  \postcode{98052}
  }

\renewcommand{\shortauthors}{Rognon, et al.}

\begin{abstract}
Social touch is essential for our social interactions, communication, and well-being. It has been shown to reduce anxiety and loneliness; and is a key channel to transmit emotions for which words are not sufficient, such as love, sympathy, reassurance, etc. However, direct physical contact is not always possible due to being remotely located, interacting in a virtual environment, or as a result of a health issue. Mediated social touch enables physical interactions, despite the distance, by transmitting the haptic cues that constitute social touch through devices. As this technology is fairly new, the users’ needs and their expectations on a device design and its features are unclear, as well as who would use this technology, and in which conditions. To better understand these aspects of the mediated interaction, we conducted an online survey on 258 respondents located in the USA. Results give insights on the type of interactions and device features that the US population would like to use. 
\end{abstract}


\keywords{Mediated Social Touch, Hardware and Software that Enable Social Touch Interactions, Haptic Technology}

\maketitle

\section{Introduction}
Social touch is a physical interaction that expresses an in-tent between two or more social agents. Typical examples of social touch include shaking hands with colleagues for greeting, hugging family members for comfort, or patting a friend’s shoulder for congratulating. Social touch is essential for our development, well-being, communication and social interactions \cite{jones1985naturalistic, field2010touch, van2015social, huisman2017social}. For example, social touch reduces anxiety and loneliness, and is a key channel to transmit emotions for which words are not sufficient such as love, sympathy, reassurance, etc. \cite{field2010touch, van2015social, huisman2017social}. In addition to the need for physical contact, people also perceive some social touches, such as handshakes and hugs, as part of the social custom when interacting with other individuals. These physical interactions are key components of a culture. 

However, in some contexts, direct physical contact is not possible such as when people are remotely located, interacting in a virtual environment, experiencing health issues, or due to governmental restrictions. Nevertheless, it is crucial to support these physical interactions between individuals to create and strengthen interpersonal bonds \cite{cascio2019social, huisman2017social}. The implications of reduced physical contact can be clearly observed worldwide during the COVID-19 pandemic. As the World Health Organization guidelines recommend avoiding physical contact, we observed the craving for physical social interactions more than ever with people finding creative ways to still interact tactilely.  However, this need for people to maintain a physical connection despite barriers of distance or safety is not unique to the COVID-19 pandemic. In fact, a new strategy has emerged over the last decade, called Mediated Social Touch (MST), that allows users to convey social touch remotely using an MST device. By using this strategy, social agents can keep this essential physical connection. Similar to direct touch, MST has been shown to increase sympathy, empathy, co-presence, and trust toward the interlocutor during communication \cite{van2015social, huisman2017social, brave2001force, bailenson2008virtual, takahashi2011improving, sallnas2010haptic}. The Midas effect — the effect that a brief touch to the hand, arm, or shoulder can positively influence attitudes towards the initiator of the social touch — has also been widely observed from MST \cite{haans2014virtual, spape2015meaning}. 

To enable MST interactions, various types of devices have been developed. They are usually classified as either wearable or non-wearable devices. Examples of wearable devices are gloves \cite{singhal2017flex}, sleeves \cite{israr2018towards, huisman2013tasst, nunez2020investigating, simons2020contact}, bracelets \cite{pezent2019tasbi, HeyBracelet, BondTouch}, jackets \cite{teh2012mobile, chung2009stress, vaucelle2009design}, and belts \cite{tsetserukou2010haptihug}. Non-wearable device examples include handles \cite{rantala2011role, Frebble}, robotic arms \cite{nakanishi2014remote}, and cellphone covers \cite{park2012couples}. While each device type has its advantages, both wearable and non-wearable devices share a common issue of being application-specific with a limited information bandwidth which prevents personalization of the tactile messages. Yet, the range of human emotional interactions we aim to carry out through social touch do indeed contain various subtle meanings and are unique to each individual and can also be situational. Expressions of social touch themselves are largely dependent on context. Current device technologies lack a holistic approach to grapple with this complexity, and instead deliver MST absent in multi-functional robustness, and may lead to frustrated users with limited potential to express themselves due to technological and/or design constraints restricted to specific contexts and messaging types. 
The demands for expressing our emotions and personalizing our interactions can be observed within the increasingly soaring catalogs of emojis that are proposed in texting applications. Yet, existing devices used to communicate mediated social touch still lack the true capability to express this level of personalization or emotional range we aim to experience. Moreover, this need for a holistic frame has often been identified in MST literature \cite{van2015social, huisman2017social, eid2015affective}. Indeed, the needs of the MST device users and their perception of this relatively recent technology are unknown. 

With this research, our objective is to begin to identify criteria which contributes to meaningful MST interactions. To reach this objective, it is important to understand what types of social touch people are missing and the contexts of these interactions. In addition, while many people now rely on video and/or voice technologies to interact with each other, it is important to identify the limitations of these technologies to communicate emotions. Furthermore, diverse MST devices have been developed, but the type of devices people would like to use and what they want to communicate through them has, to the best of our knowledge, not been studied. The form factor of any device is crucial for social acceptance and the functionalities such as touch type, with whom people want to communicate, and the capability to personalize the tactile messages are all crucial to understand. An understanding and consideration of these aspects of the device will foster meaningful user experiences and may encourage people to use MST devices more broadly. Finally, the scenarios in which people would like to use an MST device, such as during a video call or in virtual reality, also play an essential aspect in understanding the interaction. Identifying all these criteria will lead to constructing a holistic framing of the MST interaction, thereby producing results that will inform design guidelines for successful and impactful devices.

In order to identify these criteria and understand the perception of the population for such technology, we deployed an online survey during the last week of May 2020, at the beginning of a global pandemic, on the USA population. In this online survey, we were interested in answering the following four research questions: 1) What type of social touch is missed and in which context? 2) What are the limitations of current technologies to communicate emotions? 3) What type of device(s) would people like to use and to communicate what? 4) In which scenarios would people use an MST device? .

\section{Material and Methods}
The online Qualtrics survey consisted of 18 questions divided into 6 themes: the respondent’s background (5 questions), the type of social touch missed and in which context (4), the limitations of current technologies to communicate emotions (1), the type of device(s) people would like to use and to communicate what (6), in which scenarios would people use an MST device (1), and a free comment section (1) (see Appendix \ref{app1} for the full survey). The recruiting of the respondents was managed by Amazon Mechanical Turk. Respondents who were located within the USA population were eligible to participate and those who completed the entire survey were compensated. The median duration of the survey was 6.63 ± 4.13 minutes (median ± std; excluding one respondent who took 8 hours to complete the survey).

\subsection{Participant background}
\begin{figure}[!t]
	\centering
	\includegraphics[width=\columnwidth]{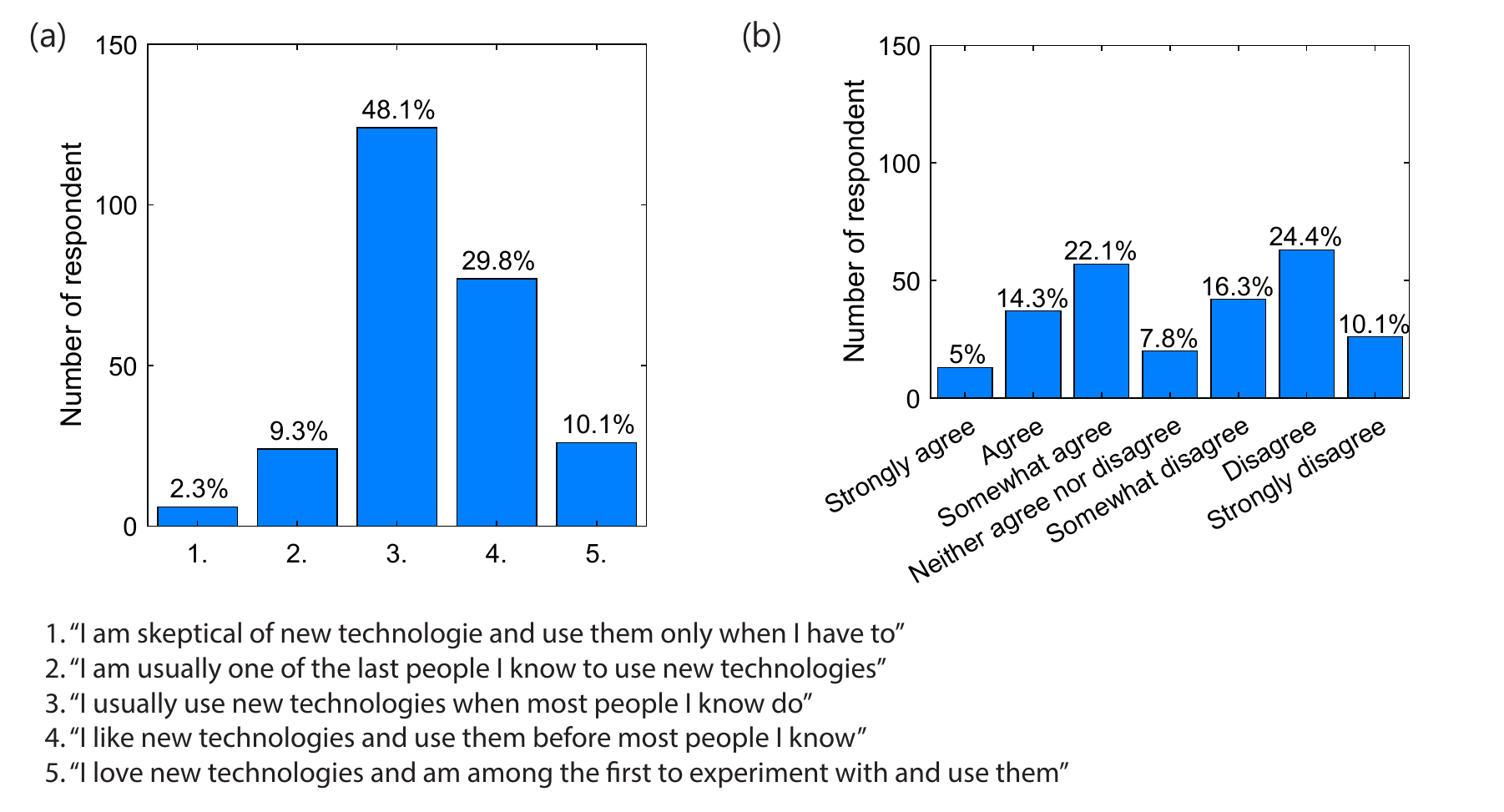}
	\caption{Results of the participant background concerning (a) their approach to new technologies, (b) their general level of willingness to engage in physical contact}
	\label{fig_partBackground}
\end{figure}

258 respondents finished the survey. 50.8 \% identified as women, 48.8 \% as men, and 0.4 \% as ``Other”. The age range spread was as follows: 1.6 \% of the respondents were within 18-24 years old, 31.9 \% within 25-34 y.o., 35 \% within 35-44 y.o., 16 \% within 45-54 y.o., 11.3 \% within 55-64 y.o., and 4.3 \% were 65 y.o. and older. The ethnicity of the respondents was mostly White (79.5 \%), then, with a much lower representation, East Asian (8.1 \%), and Black or African American (5.8 \%). The last 6.6 \% is distributed among other ethnicities. In the participant background questionnaire, we also measured the respondent’s approach to technology with the question “When it comes to technology, what best describes you?”. As shown in Fig. \ref{fig_partBackground} (a), almost half of the respondents (48.2 \%) answered “I usually use new technologies when most people do”, which is in the middle of the scale. 40.1 \% of the respondents considered themselves as an early adopter of new technologies, and 11.6 \% as not really into new technologies. The last participant background question was to assess their general level of willingness to engage in physical contact. Respondents had to answer on a 7-points Likert scale, from “Strongly agree” to “Strongly disagree”, the question “I consider myself a touchy-feely person”. Results are shown in Fig. \ref{fig_partBackground} (b). Respondents' perception is well spread with 41.4 \% considering themselves to be a touchy-feely person at a certain level, while 50.8 \% not.

\subsection{Social touch missed}
Before looking for the design of a mediated social touch device, it is important to understand what type of social touch people would like to communicate and in what context. To gather data on this aspect, respondents had to fill the blanks of the sentence “In general, I am missing physical touch such as \_[touch]\_ with my \_[person]\_ \_[amount]\_” and could choose one option for each blank (\_[touch]\_, \_[person]\_ and \_[amount]\_) from a dropdown list. Examples of options for the \_[touch]\_ were: asking for attention, holding their arm, holding hands. Examples of options for the \_[person]\_ were: partner(s), parent(s), child(ren), sibling(s)). The complete list of options can be found in Appendix \ref{app1}. The options for the \_[amount]\_ were "a slight amount" that weight 1, "a moderate amount" 2, "a lot" 3, or "an extreme amount" 4. One example of sentence is “In general, I am missing physical touch such as hugging with my friend a lot.”. Respondents could create up to three sentences. In addition, as a separate question for each sentence they created, respondents could select from a list of the emotions they wanted to communicate in each interaction. Examples of options for the emotions were affection, love, comfort, support, and they could type other emotions under the “Other” option. Respondents were not limited in the number of emotions they could select, as usually more than one emotion is felt in a situation. 

\subsection{Emotions missed in current technology}
The second section was aimed at identifying emotions that people feel unable to communicate effectively with current technologies.  For this, the respondents had to answer the following question: ``When using existing communication technologies (phone call, video call, texting, virtual reality, social media, etc.), which emotion(s) do you have difficulty communicating?'' The list of proposed emotions was the same as for the previous question including the “Other” option where they could type their own response. They could select multiple choices or the answer “No difficulties”.

\subsection{Desired social touch by technology}
Once we understood what emotions people would like to communicate, the survey asks about the types of devices respondents would like to use as well as types of messages they would like to transmit. The first two questions focused on the type of device respondents would be willing to use. The two questions were: “Without considering any technological limitations, which of the following wearable / non-wearable devices would you be willing to use to communicate touch?”. Examples of wearable options include glove, bracelet, ring, necklace. Examples of options for the non-wearable devices were cell phone, tablet, stuffed animal. For both questions, they could choose multiple options, including a free entry option “Other”, or “None of these”. The next four questions of the survey asked about the content of the tactile message respondents would transmit. Respondents had to answer the four following questions: “Without considering any technological limitations, with whom / which touch(es) / which gesture(s) / what kind of emotion(s) would you like to communicate using a device that can transmit touch?”, with again the possibility to select multiple options. For definition, touch refers to social touch such as handshake, fist bump, hugging, cuddling, etc., while the gestures are the movements that compose these social touches such as holding, pressing, poking, etc. The lists of options for the person, touch, and emotions were the same as mentioned in the first question with the added input text entry “Other” and the options “I would not use a device to communicate touch messages with anyone”, “I would not like to communicate touch using a device”, and “I would not like to communicate emotion using a device” respectively. Examples of gesture options were poking, squeezing, stroking, pinching, “Other”, and “I would not like to communicate gestures using a device”.

\subsection{Scenarios of social touch}
Finally, we were interested to know the scenarios in which respondents would be likely to use an MST device. Respondents had to answer the question: “Without considering any technological limitations, how likely are you to communicate touch in the following scenarios?” In real life, during a video call, during a voice-only call, while texting, in virtual reality, with a standalone device. They had to rate each scenario on a 5-points Likert scale from 1 - “Not at all” to 5 - “Absolutely”.
At the conclusion, respondents were provided with an open-response section where they could express themselves and give comments about any parts of the survey.

\section{Results}
\subsection{What types of social touch are people missing and in which context?}
\label{sec_missing}

\begin{figure}[!t]
	\centering
	\includegraphics[width=\columnwidth]{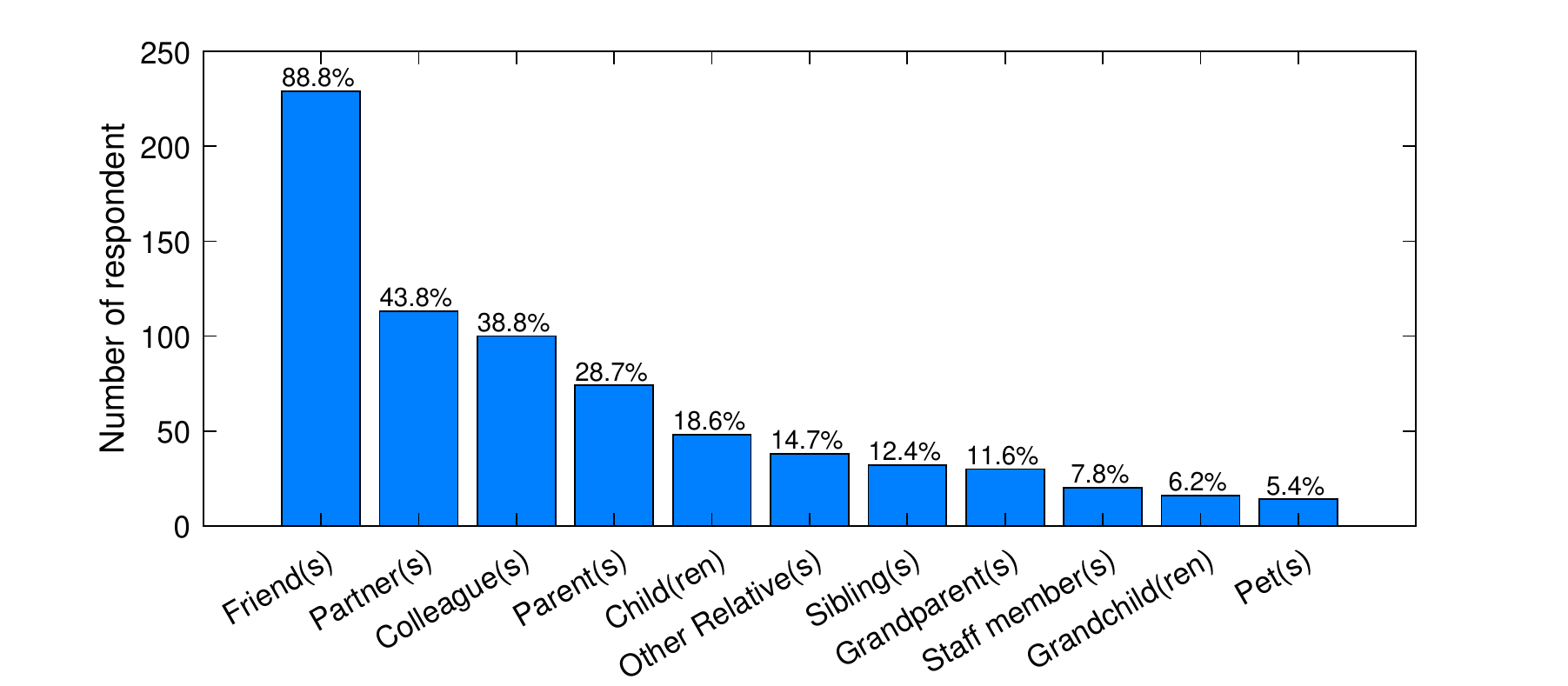}
	\caption{Results for the missed social touch interactions by relationship. The numbers above the bar plot represent the percentage of respondents that selected this option}
	\label{fig_peopleMissed}
\end{figure}

\begin{figure*}[!t]
	\centering
	\includegraphics[width=\textwidth]{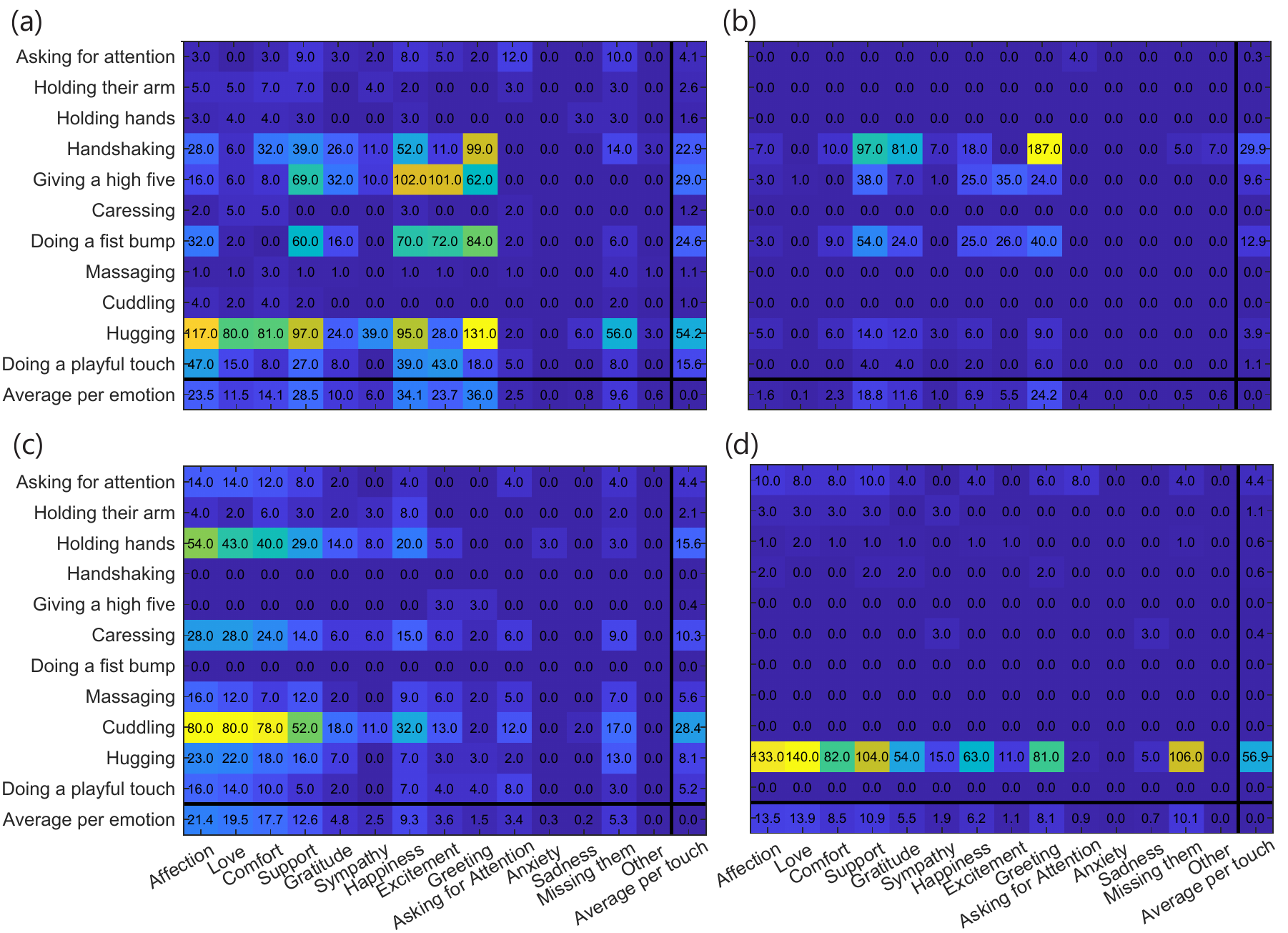}
	\caption{Heatmap representing the weighted occurrences of each social touch interaction scenarios that are missed by the respondents with (a) their friends, (b) their colleagues, (c) their partner, and (d) their parents}
	\label{fig_heatmap}
\end{figure*}

Figure \ref{fig_peopleMissed} shows the results of the missed social touch interactions by relationship. Friend(s) are the most missed people with 229 respondents (88.8 \%) selecting this option, then partner(s) (43.8 \%), colleague(s) (38.8 \%) and parent(s) (28.7 \%). It is worth noting that since this survey was conducted during a time of high social distancing, regularly social touches with friends are reduced more than that of partners, which may have impacted upon these results. In addition, the authors expect a bias for grandparent(s) (11.6 \%), grandchild(ren) (6.2 \%) and pet(s) (5.4 \%) as we do not expect all respondents to have these social agents in their life.
Results for the type of social touch that are missed and their context for the four most missed people (friend(s), colleague(s), partner(s), and parent(s)) are displayed with heatmaps on the Fig. \ref{fig_heatmap} (a), (b), (c), and (d), respectively. The heatmaps display the match between the social touches in the y-axis and the emotions in the x-axis. Numbers in the table are the occurrences of each scenario, weighted by the amount the respondents were missing the interaction (“a slight amount"= 1, "a moderate amount" = 2, "a lot" = 3, or "an extreme amount" = 4). For example, a respondent completed the following fill-in-the-blank sentence: “In general, I am missing physical touch such as hugging with my friend a lot.” Then they selected affection and greeting on the follow-up question: “what kinds of emotion(s) would you like to communicate?” This result is displayed in the heatmap for friend(s) (Fig. \ref{fig_heatmap} (a)). The combinations of hugging and each the affection and greeting emotions is represented as three occurrences, as the respondents selected “a lot” as the amount they were missing this interaction. The averages per social touch are shown in the last column and the averages per emotion in the last row. 

Fig. \ref{fig_heatmap} (a) shows the results for the friend(s). Respondents indicated missing a large variety of social touch types and emotions with their friends. Hugging is the most popular social touch interaction among respondents and is used to communicate a variety of different emotions. The other most popular social touches aim to greet other people using high fives, fist bumps, handshakes, etc. but they are also used to express a broad variety of emotions beyond initial greetings such as affection, support, happiness, love, comfort, excitement, etc.

The most popular interaction among colleague(s) is the handshake used for greeting (see Fig. \ref{fig_heatmap} (b)). The 187 score for handshake is the highest score of any scenario in this context, with next highest scoring touches, fist bump (40) and high five (24), being a fraction of this score. The other popular emotions that are missed when interacting with colleagues are support and gratitude both expressed with the handshake gesture as well.

Respondents miss more intimate social touches with their partners, and the variety of touch is spread (see Fig. \ref{fig_heatmap} (c)) with intimate social touches having the highest scores. The most popular intimate social touches are cuddling, which transmit a lot of affectionate emotions, followed by holding hands and caressing to transmit the same kind of emotions. On the contrary, greeting partners using social touch is the least popular. Social touches and emotions are both more distributed than the interactions with colleagues and parents (see Fig. \ref{fig_heatmap} (b) and (d) respectively). 

Hugging is by far the most popular social touch that respondents miss with their parents (see Fig. \ref{fig_heatmap} (d)). Respondents report using hugs to transmit a large range of emotions, mostly positive, such as love, affection and sup-port. But they also would like to transmit “missing them”, which is not a popular emotion among the three other social agents. 

The proposed negative emotions, sadness and anxiety, were rarely selected. It appears that people are not missing communicating these emotions in the context of social touch. 
Interestingly, four respondents entered “respect” on the “Other” option mostly when performing handshakes and when interacting with friends. Other free entries were friendliness, leaving, and social custom. Again, we can observe that handshakes transmit a large variety of emotions but are also primordial in social convention.

\subsection{What are the limitations of current technologies to communicate emotions?}
\label{sec_limitation}

\begin{figure}[!t]
	\centering
	\includegraphics[width=\columnwidth]{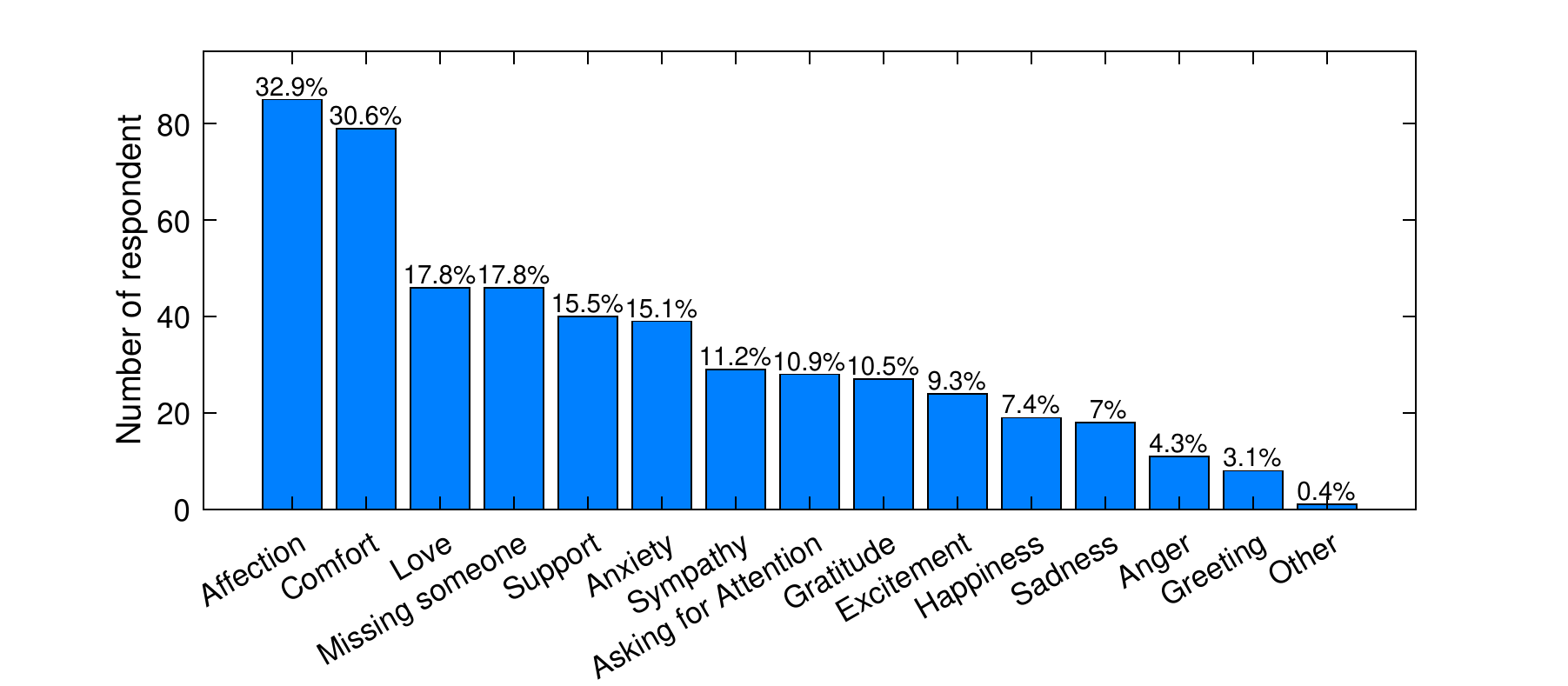}
	\caption{Results for the emotions that respondents found the most limited to communicate with current technologies}
	\label{fig_limTech}
\end{figure}

Concerning feeling limited by current technology to communicate emotions, 64.3 \% of respondents felt that they have difficulties communicating some emotions with current technologies and on average they selected three emotions each. The emotions that were difficult to communicate are quite similar to the one's respondents are missing (see Fig. \ref{fig_limTech}) with high rating for affection (selected by 32.9 \% of the respondents) and comfort (30.6 \%). There is also a relatively high rating for missing someone (17.8 \%, the third most cited emotions together with love), and anxiety (15.1 \%, 6$^{th}$ most cited emotion). However, on the opposite to the emotions people are missing, greeting has not been often selected (3.1 \%).

\subsection{What type of device(s) would people like to use and to communicate what?}
\label{sub_devices}
77.1 \% of the respondents selected some types of devices for the questions on wearable and 79.8 \% for non-wearable devices. 14.0 \% of the whole respondents selected “None of these” for both questions, which means that they would not use any listed devices nor choose the “Other” option. 

\begin{figure}[!t]
	\centering
	\includegraphics[width=\columnwidth]{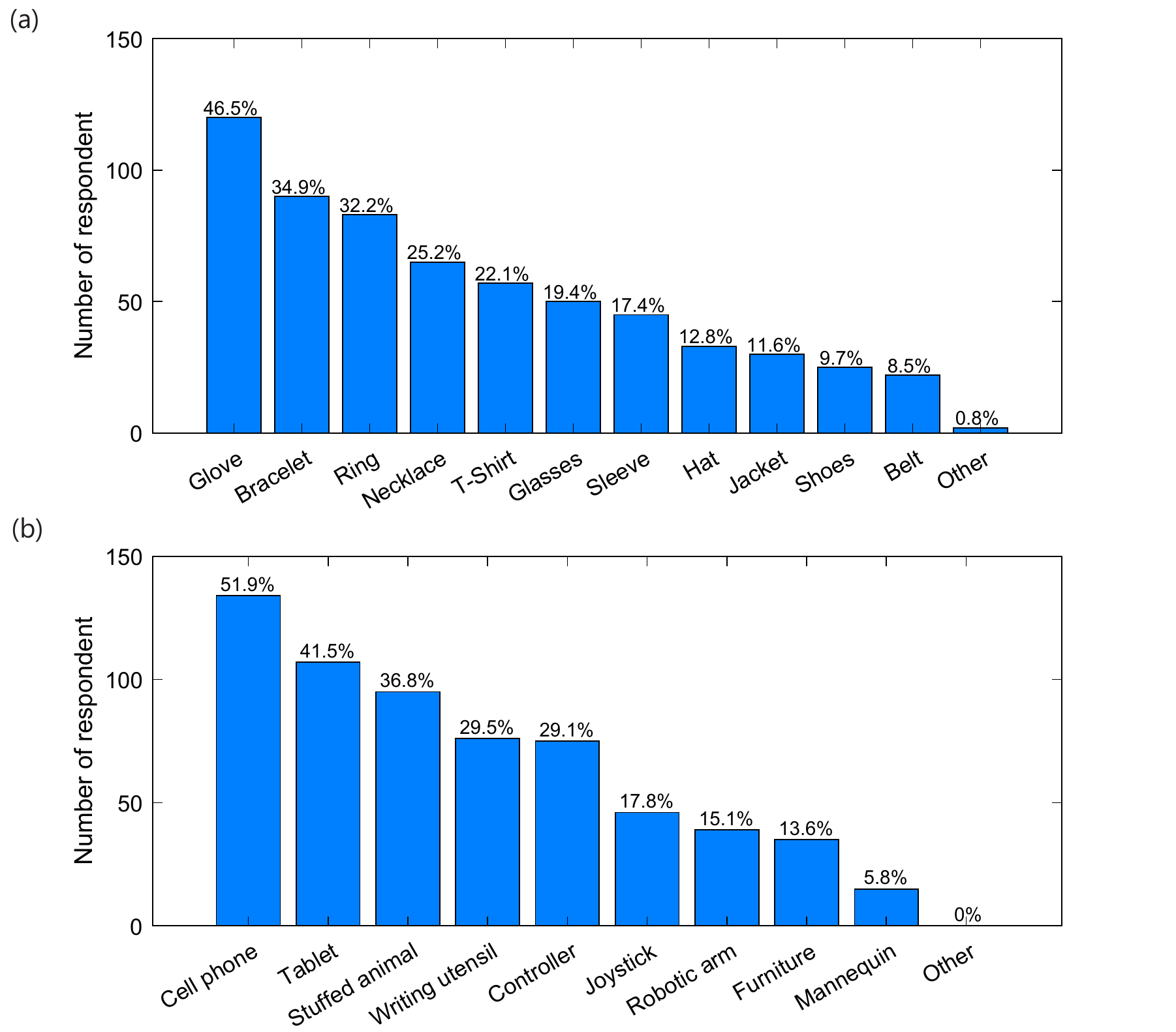}
	\caption{Results for the selected (a) wearable and (b) non-wearable devices that respondents would use to communicate social touch}
	\label{fig_devices}
\end{figure}

Concerning the type of wearable devices; the glove is by far the most popular option with 120 respondents selecting this option (46.5 \%) (see Fig. \ref{fig_devices} (a)), followed by bracelet (34.9 \%) and ring (32.2 \%). We can also see that these top three devices are located in the hand region. In addition, the most popular wearable devices are small and discreet. With necklace being the fourth most selected wearable device (25.2 \%), $\frac{3}{4}$ of the most popular options are jewelry. 

Concerning the non-wearable types of devices; the most popular devices are the cell phone (51.9 \%), followed by the tablet (41.5 \%) (see Fig. \ref{fig_devices} (b)). The high rating for the stuffed animal (36.2 \%) is interesting, and we can make a hypothesis that respondents would feel comfortable communicating emotions through a comforting object. Human-like devices such as robotic arm (15.1 \%) or mannequin (5.8 \%) are not popular. Maybe these induce an uncanny valley effect — which is, as a very short definition, an eerie, or uncomfortable, feeling that some people may encounter when looking at an entity that is almost, but not quite, human \cite{mori2012uncanny}. We can also see that the most selected options are the most portable devices and the top two are familiar technological objects. There is no significant preference for either wearable or non-wearable devices among respondents.

\begin{figure*}[!t]
	\centering
	\includegraphics[width=\textwidth]{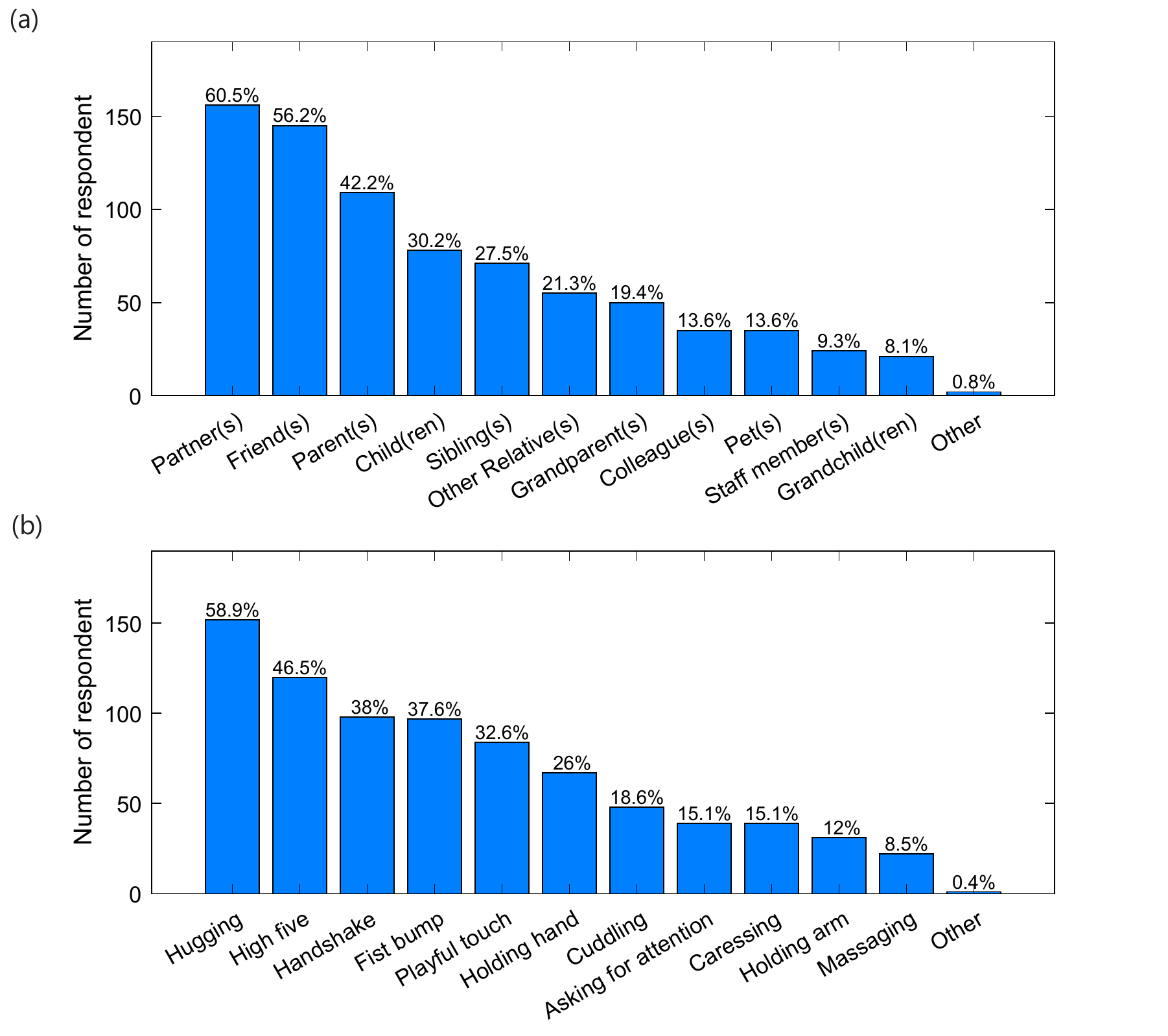}
	\caption{Content of the mediated social touch interaction that respondents would like to have. (a) The social relationships and (b) the types of social touch}
	\label{fig_convey1}
\end{figure*}

Of the 88.0 \% of respondents that would like to communicate with someone using MST, the partner(s) was the most often selected (60.5 \%, see Fig. \ref{fig_convey1} (a)). Then the ranking is the same as with who they are missing physical interactions, with friend(s) (56.2 \%) and parent(s) (42.3 \%), but except for theirs colleague(s) (13.6 \%). It seems that respondents would like to use MST mostly to communicate with loved ones.  

Of the 86.0 \% of respondents that selected at least one social touch that they would like to communicate, hugging (58.9 \%) was the most often selected, which corroborate with the type of social touch they are missing (see Fig. \ref{fig_convey1} (b)). The raking of these social touches is also very similar as the ones they are missing.

\begin{figure*}[!t]
	\centering
	\includegraphics[width=\textwidth]{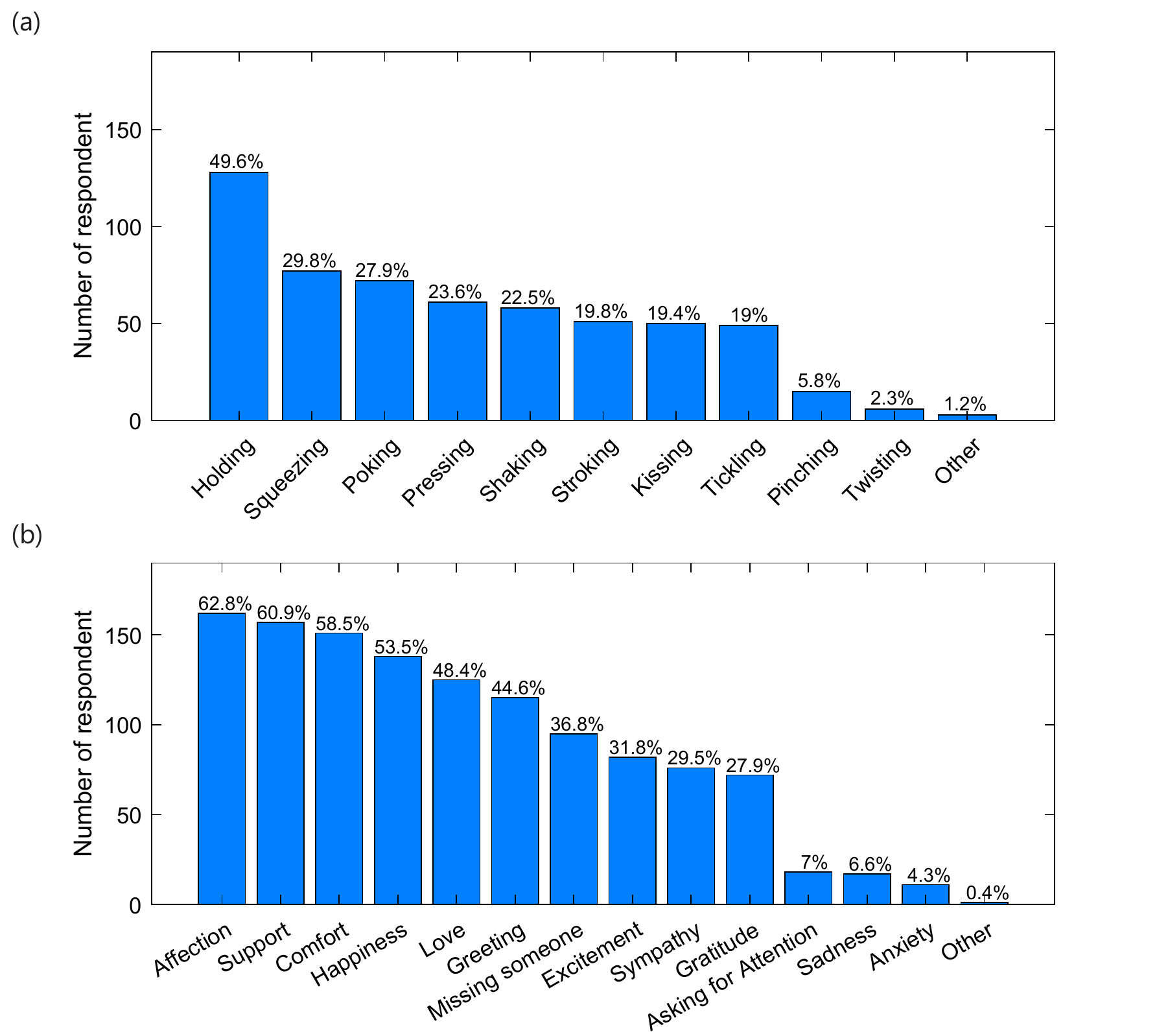}
	\caption{Content of the mediated social touch interaction that respondents would like to have. (a) The gestures and (b) the emotions they would like to convey}
	\label{fig_convey2}
\end{figure*}

Concerning the gestures, 79.5 \% of the respondents selected some options. This is the lower percentage of selection. Maybe it was hard for some respondents to envision such gestures. Holding is the most popular one (49.6 \%), which we hypotheses may be correlated with hugging (see Fig. \ref{fig_convey2} (a)). Pinching (5.8 \%) and twisting (2.3 \%) are the least popular. One explanation could be the painful image that these gestures may trigger.

And finally, Figure \ref{fig_convey2} (b) show the results for the question asking for the emotions respondents would like to transmit using an MST device. The ranking is very similar to the emotions they are missing, and the ones they have difficulties to communicate using current technologies with the quartet affection (62.8 \%) - support (60.9 \%) - comfort (58.5 \%) - love (48.5 \%) being often selected. 86.5 \% of the respondents selected some emotions. They would like to be able to transmit a lot of different emotions with the MST device as they selected on average 4.7 emotions. Respondents have no wish to communicate negative emotions such as sadness (6.6 \%) and anxiety (4.3 \%), even if they stated that they had difficulties to transmit anxiety when using current technologies (see subsection \ref{sec_limitation}).

8.9 \% of the respondents answered “None of these” to the four questions. In fact, 7.4 \% of the respondents selected “None of these” to both questions about the device type and to the four questions about the content of the communication.

\subsection{In which scenarios would people use an MST device?}
\begin{figure}[!t]
	\centering
	\includegraphics[width=\columnwidth]{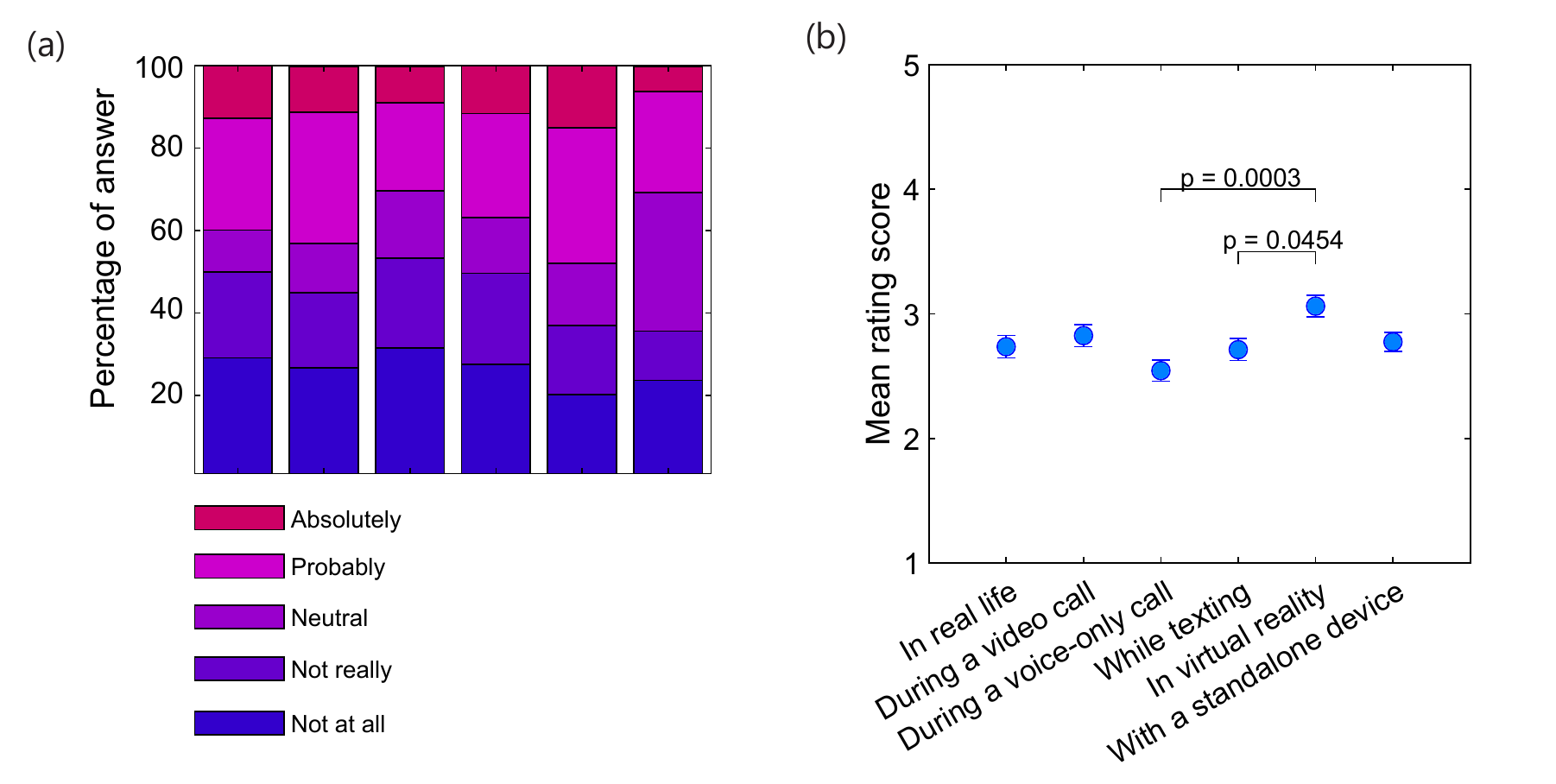}
	\caption{Scenarios in which respondents would like to use an MST device. (a) Distribution of the answers, (b) mean rating score for each interaction scenario}
	\label{fig_scenario}
\end{figure}

Figure \ref{fig_scenario} (a) shows the distribution of the answer from “Not at all” to “Absolutely” for each scenario. We can see that the “In virtual reality” and “During a video-call” scenarios have a higher percentage of high rating such as “Absolutely” (15.1 \% and 11.2 \%) and “Probably” (32.9 \% and 31.8 \%). On the opposite, “During a voice-only call” and “while texting” have a larger percentage of low rating such as “Not at all” (31.5 \% and 27.5 \%) and “Not really” (21.8 \% and 22.1 \%).
When we compute the mean and standard error, we can see a significant difference of rating between the “In virtual reality” scenario and “During voice-only call” and between “In virtual reality” and “While texting” (see Fig. \ref{fig_scenario} (b)). However, as the difference in mean rating score is quite small, we had a look at the size effect of these distributions by calculating the Cohen’s d coefficient. As a reminder, a higher coefficient means a larger size effect. 0.2 is considered as a small effect, 0.5 as a medium effect, and 2 as a ‘huge’ effect. We can see here that the coefficients are quite small for both tests (0.38 between “During a voice-only call” and “In virtual reality” and 0.25 between “While texting” and “In virtual reality”. The size effect is probably small as results have large variances, so we had a closer look at the responses of individuals. 
We observed that 76.7 \% of respondents answered “Probably” or “Absolutely” to at least one scenario and 32.6 \% of the whole respondents answered “Absolutely” to at least one scenario. We hypothesis that these are people that would be inclined to try an MST device. On the opposite, only 11.2 \% put “Not really” or “Not at all” for every scenario and 7.4 \% “Not at all” for every scenario. Half of these respondents (4.3 \% and 3.5 \% of the whole respondents respectively) are the same as the ones that selected only “None of these” to all the questions about the type of device (see subsection \ref{sub_devices}).

\section{Discussion}
Results of the online survey give insights on the type of interaction people are missing and would like to transmit using an MST device. People are primarily missing their friend(s), partner(s), colleague(s), and parent(s) but they would not use the same type of social touch to express their emotions with these different types of people. Social touch to greet, such as handshake, high five, fist bump, and hugging, would be used to interact with friend(s) and colleague(s). These results point out the importance of a physical contact when starting an interaction with someone. However, the emotions that people want to transmit with these social touches are not only to greet them but also to transmit various types of emotions such as affection, comfort, and even happiness and excitement. Social touches used with partner(s) are more intimate as expected but also more varied. On the opposite, almost only hugging was selected to interact with the parent(s). These results demonstrate that the type of social touch that people want to communicate is strongly dependent on the relationship between the social agents but also that the same social touch can have different meaning depending on the nature of the interactions. Therefore, it is important to enable the transmission of a large range of social touches and allow a significant degree of personalization during MST interactions. 

Respondents answered that they felt limited when communicating emotions using current technologies to transmit similar emotions that the ones they are missing (except for missing someone and anxiety). These results make sense as usually people are not missing negative emotions, but they still want to be able to share their troubles. Respondents did not feel limited to greet someone even though they are greatly missing this social interaction. This shows again that a tactile contact when meeting someone is important. Using current technologies, respondents are still able to greet their interlocutors, but because of technological limitations, users cannot experience this physical interaction. Affection and comfort are emotions that are hard to communicate through speech, as opposed to happiness, asking for attention or excitement that were rated lower. These findings indicate that MST could play an essential role in interactions by enabling the communication of these emotions.

Knowing the need of the population for communicating through social touch, the second part of the online survey was investigating respondents’ perceptions of MST devices. It is encouraging to observe that a large part of the respondents showed interest in this new technology. Indeed, the majority of the respondents selected multiple options, and only 7.4 \% of the participants showed no interest by selecting “None of these” to all the questions regarding the MST device. This conclusion was also back up by the free comments section (see a later paragraph). Concerning the type of MST devices, respondents showed preferences for small, discreet, and familiar objects such as cell phones, tablet, ring, or bracelet with the glove being an exception but making sense as a large part of the interaction are made through the hands. There is no significant preference for either wearable or non-wearable devices among respondents. 

Results also showed that while they are missing interacting through social touch with their colleagues, respondents are not willing to use an MST device to communicate with them. Indeed, they would rather use an MST device to communicate with their loved ones such as their partner(s), parent(s), or friend(s). This result illustrates the intimate essence of touch. Even if we usually use social touch to interact with a lot of people, vendors, staff members etc., it is important that they happen in a controlled environment and that we trust the person we are interacting with.
Respondents would like to be able to use a device to transmit a large variety of social touch, which supports the claim that the device should be multipotent. These social touches should also be able to convey a large range of emotions, thus the need for personalization.

The last section of the survey investigated the interactions scenarios. Results indicated the importance of the multisensorial experience of a social touch interaction, meaning that respondents would like to communicate through MST while also having visual and auditory feedback. Indeed, most of the respondents would prefer to use an MST device in virtual reality or during a video call instead of while texting or during a voice call.

While analyzing the data, we tried to draw a profile of preferences regarding the background of the user such as for example the type of social touch that people within a certain age range would like to transmit or the type of devices that respondents who consider themselves as not “touchy-feely" would feel comfortable with. Unfortunately, the cross-correlation analysis with the respondents’ background did not show any statistical significance. Nevertheless, we observed that the 6.2 \% of the respondents that selected “No difficulties” for the question about the limitations of current technologies, “None of these” to both questions about the device type and to the four questions about the content of the communication, are all considering themselves non-touchy people. Otherwise, they have no specific background.
However, using the free comments section and by looking at the whole survey results, we observed two profiles of respondents: we called them “the skeptics” and “the optimists”. Indeed, we noticed that a percentage of respondents that seem skeptic about the proposed concepts of MST technologies by selecting the “None of these” option to all the six questions regarding to the MST device, “Not at all” for every scenario, or/and by stating it in the free section comments. We identified 37 (14.3 \%) skeptical respondents. From the free comment section, it appeared that these skeptic respondents have three main concerns: first, that communicating with MST would feel “weird” and “creepy”, secondly that MST could never replace real touch or third, they just do not need such technology. We can see that the two first concerns focus primarily on the tactile sensations of the interface with the skin, the rendering of the haptic primitives by the actuators and the realism of the sensations, and on the feeling of co-presence and believability. These concerns can be addressed by designing pleasant interfaces, by studying the emotion decoding and recognition rate of the tactile message, the emotional reception of the tactile messages, and the interaction immersion and feeling of presence. We can also observe the similitude between this population behavior and the ambivalent public reaction to the introduction of the telephone \cite{fischer1992america}.
We also identified 99 (38.4 \%) optimistic respondents by using results of the free comment section, their scores at the scenario likeliness questions and by proofreading their answers through the whole survey. Their main outputs were that they found the MST concepts ``very interesting'' and ``innovative'', that ``they would like to know more about this technology'', and that ``this survey made them think about these concepts''. Some respondents even highlighted that they could feel the negative impact of physical distancing on their life. However, even if they are optimist about such technology, we observed that respondents had difficulties to imagine such devices. On top of the free comment answers, we observed that 13.9 \% of the whole poll of respondents selected “None of them” for both questions about the types of devices but only 9.0 \% selected “None of these” to the four questions about the content of the tactile messages.
The factor that plays a primordial role in the development of MST device is the novelty barrier. Indeed, we observed in this survey that most of the respondents felt positive about such technology but not many were familiar with this concept. This novelty barrier needs to be overcome. To do this, device designers should take social norms and trends into account. For example, a device could be hidden under the shirt or on the opposite very visible on the wrist or on the neck (such as a jewelry), to go along with the current social norms.
  
These are the recommendations that can be drawn out of the results of this survey. However, it is important to keep in mind that respondents were located in the USA only and in addition, unfortunately, this survey suffer from a lack of diversity in ethnicity with mostly white respondents. We know that social touch interactions vary considerably between cultures. Therefore, the design of the MST device should take into account the target population or allowing to be extremely versatile. 
A second limitation is the influence of the respondents being in a pandemic for the past couple of months when the data were collected. We do not think that this has a significant influence on the results except that respondents may be craving more social touch interactions.  

In conclusion, the main take home messages from this research are that social touch can vary regarding the interlocutors and it can have various meanings depending on the context of the interaction. Concerning the device design, respondents felt more comfortable to communicate with loved ones using MST devices that have small and familiar form factors. The MST device should allow to transmit various types of social touch that are personalizable regarding the emotions they want to convey to their interlocutor. The focus should primarily be put on conveying positive emotions. In addition, the multisensorial aspect of the social touch interaction is essential to be preserved.

\bibliographystyle{ACM-Reference-Format}
\bibliography{referencesLagunitas}

\appendix

\section{Full questionnaire}
\label{app1}
\subsection{Participant background}

Q1 What is your age?
\begin{itemize}
\item Under 18  
\item 18 to 24  
\item 25 to 34  
\item 35 to 44   
\item 45 to 54  
\item 55 to 64  
\item 65 or older  
\end{itemize}

Q2 What is your gender?
\begin{itemize}
\item Female  
\item Male  
\item Other:  \_
\item Prefer not to answer   
\end{itemize}

Q3 Which of the following categories best describes your race/ethnicity?
\begin{itemize}
\item American Indian or Alaskan Native   
\item Black, Afro-Caribbean or African American   
\item East Asian   
\item Hispanic or Latino  
\item Middle Eastern or Arab   
\item Native Hawaiian or Other Pacific Islander  
\item South Asian or Indian   
\item White   
\item Other:   \_
\item Prefer not to answer  
\end{itemize}

Q4 When it comes to technology, what best describes you?
\begin{itemize}
\item I am skeptical of new technologies and use them only when I have to 
\item I am usually one of the last people I know to use new technologies  
\item I usually use new technologies when most people I know do  
\item I like new technologies and use them before most people I know  
\item I love new technologies and am among the first to experiment with and use them 
\end{itemize} 

Q5 
Please rate your reaction to the following statement: "I consider myself a touchy-feely person."
\begin{itemize}
\item Strongly agree  
\item Agree   
\item Somewhat agree   
\item Neither agree nor disagree  
\item Somewhat disagree   
\item Disagree  
\item Strongly disagree   
\end{itemize}

\subsection{The type of social touch missed and in which context}

Q6 Fill in the blanks for the sentence below using the drop-down menu to express how you are feeling.   
You may create up to three sentences.  
Note: Do not write sentences about people you are currently living with.  
   
In general, I am missing physical touch such as \_[touch]\_ with my \_[person]\_  \_[amount]\_.   
   
\emph{Example: In general, I am missing physical touch such as \_[hugging]\_ with my \_[grandparent(s)]\_  \_[an extreme amount]\_.}   

Amount: 
\begin{itemize}
\item an extreme amount
\item a lot
\item a moderate amount
\item a slight amount  
\end{itemize}

Person:
\begin{itemize}
\item partner(s) 
\item parent(s) 
\item child(ren)  
\item sibling(s)  
\item grandparent(s) 
\item grandchild(ren)  
\item other Relative(s)  
\item friend(s)  
\item colleague(s)  
\item staff member(s)  
\item pet(s)  
\end{itemize}

Touch:
\begin{itemize}
\item asking for attention 
\item holding their arm  
\item holding hands  
\item handshaking  
\item giving a high five  
\item caressing  
\item doing a fist bump  
\item massaging  
\item cuddling  
\item hugging  
\item doing a playful touch  
\end{itemize}

Q7-9 (3x the same question regarding the respondent’s answer to Q6). 
With your \emph{SelectedAnswers}, when \emph{SelectedAnswers}, what kinds of emotion(s) do you like to communicate? 
Select all that apply:
\begin{itemize}
\item Affection 
\item Love  
\item Comfort  
\item Support  
\item Gratitude 
\item Sympathy  
\item Happiness  
\item Excitement  
\item Greeting 
\item Asking for attention  
\item Missing them  
\item Anxiety  
\item Sadness  
\item Other:  \_
\item I don't know  
\end{itemize}

\subsection{The limitations of current technologies to communicate emotions}

Q10 When using existing communication technologies (phone call, video call, texting, virtual reality, social media, etc.), which emotion(s) do you have difficulty communicating? 
Select all that apply:
\begin{itemize}
\item Affection  
\item Love  
\item Comfort  
\item Support  
\item Gratitude  
\item Sympathy  
\item Happiness  
\item Excitement 
\item Greeting 
\item Asking for Attention  
\item Missing someone  
\item Anxiety  
\item Sadness  
\item Anger  
\item Other:  \_
\item I do not have difficulty communicating emotion using existing technologies  
\end{itemize}

\subsection{The type of device(s) people would like to use and to communicate what}

Q11 Without considering any technological limitations, which of the following wearable devices would you be willing to use to communicate touch? 
Select all that apply:
\begin{itemize}
\item Glove  
\item Sleeve  
\item Glasses 
\item Shoes   
\item Ring  
\item T-shirt  
\item Hat 
\item Belt  
\item Bracelet 
\item Jacket  
\item Necklace 
\item Other:  \_
\item None of these 
\end{itemize}

Q12 Without considering any technological limitations, which of the following non-wearable devices would you be willing to use to communicate touch? 
Select all that apply:
\begin{itemize}
\item Controller  
\item Cell phone 
\item Tablet  
\item Joystick  
\item Robotic arm  
\item Mannequin  
\item Stuffed animal  
\item Furniture (chair, bed, table, lamp, etc.)  
\item Writing utensil (pen, pencil, etc.) 
\item Other:  \_
\item None of these 
\end{itemize}

Q13 Without considering any technological limitations, with whom would you use a device to communicate touch? 
Select all that apply:
\begin{itemize}
\item Partner(s) 
\item Parent(s) 
\item Child(ren)  
\item Sibling(s)  
\item Grandparent(s) 
\item Grandchild(ren)  
\item Other Relative(s)  
\item Friend(s)  
\item Colleague(s)  
\item Staff member(s)  
\item Pet(s)  
\item Other: \_
\item I would not use a device to communicate touch messages with anyone 
\end{itemize}

Q14 Without considering any technological limitations, which touch(es) would you like to communicate with using a device that can transmit touch? 
Select all that apply:
\begin{itemize}
\item Asking for attention 
\item Handshake  
\item High five  
\item Fist bump  
\item Playful touch  
\item Hugging  
\item Caressing  
\item Holding hand  
\item Holding arm  
\item Cuddling  
\item Massaging  
\item Other:  \_
\item I would not like to communicate touch using a device  
\end{itemize}

Q15 Without considering any technological limitations, which gesture(s) would you like to communicate using a device that can transmit touch? 
Select all that apply:
\begin{itemize}
\item Poking  
\item Pinching  
\item Shaking  
\item Tickling  
\item Squeezing  
\item Pressing  
\item Holding  
\item Stroking  
\item Kissing  
\item Twisting  
\item Other:  \_
\item I would not like to communicate gestures using a device  
\end{itemize}

Q16 Without considering any technological limitations, what kinds of emotion(s) would you like to communicate using a device that can transmit touch? 
Select all that apply:
\begin{itemize}
\item Affection  
\item Love  
\item Comfort  
\item Support  
\item Gratitude  
\item Sympathy  
\item Happiness  
\item Excitement  
\item Greeting  
\item Asking for attention  
\item Missing someone  
\item Anxiety  
\item Sadness  
\item Other:  \_
\item I would not like to communicate emotion using a device  
\end{itemize}

\subsection{In which scenario would people use an MST device}

Q17 Without considering any technological limitations, how likely are you to communicate touch in the following scenarios?
	5-points Likert scale: 1 -Not at all, 2 - Not really, 3 - Neutral, 4	- Probably, 5 -	Absolutely 
\begin{itemize}
\item In real life (i.e. meeting the person 6 feet apart)
\item During a video call	
\item During a voice-only call
\item While texting
\item In virtual reality
\item With a standalone device
\end{itemize}

\subsection{Free comment section}
Q18 This is the last question, please write below if you have any comments regarding this survey

\end{document}